\documentclass[twocolumn,10pt]{IEEEtran}

\usepackage{amsmath}
\usepackage{amsfonts}
\usepackage{epsfig}
\usepackage{amssymb}
\usepackage[nospace]{cite}
\usepackage{color,soul}
\usepackage{subfigure}
\usepackage{multirow}
\usepackage{rotating}
\usepackage{graphicx}
\usepackage{tabularx}
\usepackage{array}
\usepackage{graphicx,dblfloatfix}
\usepackage{blindtext}
\usepackage{ragged2e}
\usepackage{xcolor,colortbl}
\definecolor{Gray}{gray}{0.85}

\usepackage{amsthm,amssymb,amsmath,bm}
\usepackage{fancyhdr}
\usepackage[subfigure]{tocloft}
\usepackage[font={small}]{caption}
\usepackage{graphicx}
\usepackage[caption=false]{subfig}
\usepackage[utf8]{inputenc}
\usepackage[english]{babel}
\newtheorem{theorem}{Theorem}[section]

\newtheorem{lemma}[theorem]{Lemma}
\usepackage{lipsum} 
\usepackage{algorithmic}
\usepackage[Algorithm]{algorithm}
\usepackage{tabularx}
\definecolor{Gray}{gray}{0.85}
\newtheorem{remark}{Remark}
\begin{document}
	\title{Joint Information Theoretic Secrecy and \\Covert Communication in the Presence of \\an Untrusted User and Warden}

 \author{\IEEEauthorblockN{Moslem Forouzesh, Paeiz Azmi, \textit{Senior Member, IEEE,} Ali Kuhestani, \textit{Member, IEEE}, and Phee Lep Yeoh,\textit{ Member, IEEE. }} \textsuperscript{}\thanks{\noindent\textsuperscript{}
		
 M. Forouzesh, is with the department of Electrical and Computer Engineering, Tarbiat Modares University, Tehran, Iran,  (e-mail:  m.Forouzesh@modares.ac.ir.)

P. Azmi, is with the department of Electrical and Computer Engineering, Tarbiat Modares University, Tehran, Iran, (e-mail:  pazmi@modares.ac.ir).

A. Kuhestani is with the Communications and Electronics Department, Faculty of Electrical and Computer Engineering, Qom University of Technology, Qom, Iran, and also with the Department of Electrical and Computer Engineering, Tarbiat Modares University, Tehran, Iran, (e-mail: a.kuhestani@modares.ac.ir).

P. L. Yeoh is with the School of Electrical and Information Engineering,  The University of Sydney, NSW, Australia (e-mail:  phee.yeoh@sydney.edu.au)}}

{}
%
	\maketitle
	
	\begin{abstract}
		
		In this paper, we investigate joint information theoretic secrecy and covert communication in a single-input multi-output (SIMO) system where a transmitter (Alice) is communicating with two legitimate users (Bob and Carol). We consider that an untrusted user and a warden node are also present in the network attempting to attack the secure and covert communications to Bob and Carol, respectively. Specifically, Bob requires secure communications such that his messages from Alice are not decoded by the untrusted user, while Carol requires covert communications such that her messages from Alice are not detected by the warden.
		To do so, we consider that Alice transmits Carol's messages during selected time slots to hide them from the warden while also transmitting Bob's messages in each time slot contentiously.
		We formulate an optimization problem with the aim of maximizing the average rate subject to  a covert communication requirement and a secure communications constraint.
		Since the proposed optimization problem is non-convex, we utilize successive convex approximation to obtain a tractable solution. We  consider practical assumptions that Alice has imperfect knowledge of the warden's location and imperfect channel state information (CSI) of Bob and Carol.
		Our numerical examples highlight  that the imperfect CSI at Carol has a more detrimental impact on the average rate compared to imperfect CSI at Bob.\\
		
	\end{abstract}
	
	\section{Introduction}\label{Introduction}
	\IEEEPARstart{T}{he} security and privacy of wireless communications is emerging as a critical consideration for network operators due to the widespread and open nature of wireless transmissions. Generally, the security protections for wireless communications have been implemented based on well-known cryptographic key-based approaches in the higher layers of the network design \cite{h-layer}. This approach is based on assuming specific constraints on the computational capacity of a wireless eavesdropper, such that it cannot discover the secret key assigned to the legitimate users to decrypt the confidential information. Recently, information theoretic secrecy (ITS) has been introduced as a promising technique for securing wireless communication in which no complicated key-exchange procedures are imposed on the network \cite{N.Yang}.  In the pioneering work,  Wyner illustrates when an eavesdropper’s channel is a degraded version of the legitimate user's channel, the transmitter and receiver are able to achieve a positive perfect  secrecy rate \cite{Wyner}. Toward this end, several techniques have been proposed to enhance the ITS:  Transmit beamforming \cite{Khisti}, \cite{A.Khisti}, antenna selection  \cite{Yang}, \cite{Alves}, cooperative techniques   \cite{Wang}, \cite{m.for}, artificial noise aided transmission \cite{X.Zhang}--\cite{M.For}, and using power domain non orthogonal multiple access (PD-NOMA) \cite{r1}, \cite{r9}.
	In ITS, the goal is to secure the \textit{content} of the confidential message from the eavesdropper. However, in other scenarios with privacy considerations, the transmitter and receiver aim to hide the \textit{existence} of their communications from a warden, which is the so-called covert communication.

In recent years, researchers have investigated covert communication in various wireless communication scenarios such as IoT applications \cite{Z. Liu, Z. Liu_1}, unmanned aerial vehicle (UAV) networks \cite{H. Wang, X. Zhou}, cooperative relaying networks \cite{J.Wang}-\cite{Forouzesh&Kuhestani},  device-to-device (D2D) communications in 5G \cite{Y. Jiang}, and IEEE 802.11 Wi-Fi networks \cite{W. Kim}. 
In  \cite{X. Zhou}, covert communication was considered in the presence of a UAV with location uncertainty of terrestrial nodes.
In \cite{Z. Liu} and \cite{Z. Liu_1} the authors investigated covert communication in an IoT network and showed that the presence of interferences from other devices can be harnessed to support covert communication.
In \cite{J.Wang}, the authors investigated covert communication in the presence of an amplify-and-forward relay under the assumption of channel uncertainty. 
Greedy relaying was investigated in \cite{Hu} in which the relay opportunistically transmits its own information to the destination covertly besides retransmitting the source's message. In \cite{W. Kim}, a covert jamming attack was investigated which is an insetting attack in IEEE 802.11 wireless LANs.  The aim of this attack is to destroy the data and defraud the transmitter by injecting a covert jamming signal \cite{W. Kim}.

Most previous works have assumed that perfect channel state information (CSI) is accessible. However,
in realistic scenarios, it is challenging to acquire the CSI of legitimate nodes without channel estimation error. This is because imperfect events like feedback delay, limited training power and duration, and
low-rate feedback \cite{m.for} impact on the channel estimation
procedure. To this end, the authors in \cite{MFICDI} studied covert communications with imperfect knowledge of the warden's channel distribution
 while perfect CSI of the legitimate user is still available. The idea of employing an uninformed jammer was proposed in \cite{jammer}, where the source can transmit data covertly to the destination in the presence of an adversary. Recently, the authors in \cite{Forouzesh} studied and compared the performance of ITS and covert communication for a single wiretap channel with the aim of maximizing the secrecy or covert rate.

	In this paper, we consider the joint ITS and covert communication requirements in a single-input multi-output (SIMO) network where two legitimate users (Bob and Carol) request two different communication scenarios from the transmitter Alice which is a novel system model and has not been considered before. Furthermore, we consider that there are two adversary nodes, an untrusted user and a warden node, present in the network performing ITS and covert communications attacks, respectively. In this system model, Bob needs to receive his message securely, while Carol needs to receive her message covertly.	For secure transmission, our aim is to prevent the untrusted user from decoding Bob's message from Alice.
	Additionally, for covert communication, our goal is to avoid the warden from detecting the presence of Carol's message from Alice. To achieve this, we consider that Alice transmits Carol's messages during selected time slots to hide them from the warden, while she transmits Bob's messages in each time slot contentiously. Different from previous works that relied on high powered external jammers for interference, our proposed joint transmission model applies Bob's data signal as interference at Willie to support the covert requirements of Carol, while Carol's data signal is interference at the untrusted user to support the ITS requirements of Bob. Based on this approach, we formulate an optimization problem with the aim of maximizing the average rate subject to the covert and secure communication constraints, i.e., preventing the detection of communications and	extraction of data by the warden and untrusted user, respectively.  Since the proposed optimization problem is non-convex, its solution is intractable. As such, we adopt successive convex approximation to convexify the objective function and obtain a tractable solution.  We also consider the practical scenario, where the CSI of the users and the location of the warden are not perfectly known.  Finally, numerical examples and discussions are	provided to highlight joint ITS and covert design insights. Specifically, we confirm that the joint secure and covert communications can be successfully achieved by our proposed transmission scheme. Furthermore, we observe that the imperfect CSI of Carol has a more negative impact on the average rate compared to Bob.

 \section{System Model}\label{Sytem Model}
\begin{figure}[h]
	\begin{center}
		\includegraphics[width=3.5in,height=1.1in]{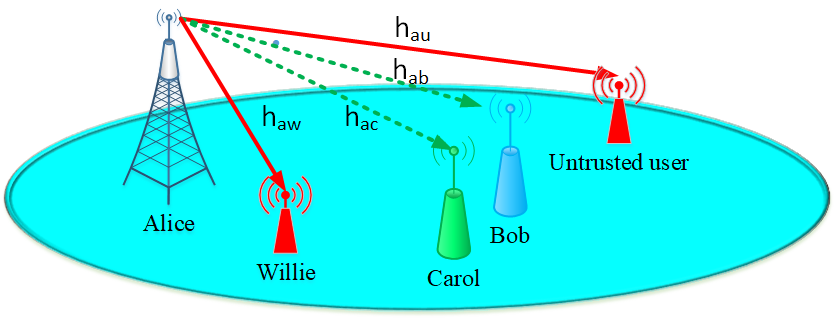}
		\caption{System model of joint secure and covert transmission.}
		\label{Sys}
	\end{center}
\end{figure}

We consider the system model shown in Fig. \ref{Sys}, which consists of one transmitter (Alice), two legitimate users (Carol and Bob), one untrusted user, and one warden. This untrusted user and warden scenario may arise in large-scale distributed systems where the trustworthiness and transparency of all users in the network is difficult to guarantee and therefore the transmitter Alice will need to adapt her communications protocol based on the requirements of the legitimate users and potential adversary users identified by the network operator. 
The distance between Alice and Bob, Alice and Carol, Alice and untrusted user, and Alice and warden are defined as $d_{ab}$, $d_{ac}$, $d_{au}$ and $d_{aw}$, respectively.
The channel fading coefficients between Alice and Bob, Alice and Carol,  Alice and untrusted user, Alice and warden are $h_{ab}$, $h_{ac}$, $h_{au}$ and  $h_{aw}$, respectively, and these channels  have circularly symmetric complex Gaussian distribution with zero mean and unit variance. We assume all the channel coefficients remain constant within one frame and change  from one frame to another independently.

 Alice  transmits confidential messages to  Carol and Bob, where one user (Bob) requires secure communications to protect against the untrusted user and another user (Carol) requires covert communications to avoid detection by the warden. 
 Hence, Alice employs a joint ITS and covert communication approach to transmit data to Bob and Carol, respectively.
  In our proposed approach, Alice transmits Carol's messages according to a predetermined set of indexes for the covert communication time slots, while she transmits Bob's messages in each data transmission time slot contentiously.
  In the considered system model, we assume that Alice knows the location and CSI of Bob, Carol and the untrusted user whereas Alice only knows the location of the warden with no CSI information. This is because we assume the untrusted user is an active user that knows the codebook of the communication network to decode the transmissions from Alice whereas the warden is a passive user that does not participate in any communications. \cite{jammer}.
  
We consider a discrete-time channel with $Q$ time slots, each having a length of $n$ symbols, hence, the transmit signals to Carol and Bob  in one time slot are ${{\boldsymbol {x}}_c} = \left[ {x_c^1,x_c^2,...,x_c^n} \right]$ and ${{\boldsymbol{x}}_b} = \left[ {x_b^1,x_b^2,...,x_b^n} \right]$, respectively. Note that  Alice transmits  $\textbf{x}_b$ continuously while she only transmits $\textbf{x}_c$ to Carol during selected covert communication time slots.  In the next section, we investigate two main cases:
 1) Only Carol knows the covert communication time slot indexes, 2) both Carol and Bob know the covert communication time slot indexes.

\section{ Proposed Joint Optimization of ITS and Covert Transmission Rate}\label{Only Carol}
In the covert communication literature, the covert strategy (index of data transmission slot) is encoded as a secret of sufficient length to be shared between Alice and Carol \cite{J.Wang},  \cite{Bash}, \cite{Shahzad}, which is unknown to the warden. In this section, we will first consider that Bob does not know Carol's covert strategy i.e., he does not have access to  Alice and Carol's pre-shared secret. In the following, we analyze the proposed system model based on this assumption.

\subsection{Information Theoretic Security Requirement}\label{Physical layer security}

The received vector at node $m$ (Bob, Carol, untrusted user, and warden) is given by:
\begin{align}
{\boldsymbol{y}_m} = \left\{ {\begin{array}{*{20}{l}}
	{\frac{{\sqrt {{p_{ab}}} {h_{am}}{\boldsymbol{x}_b}}}{{d_{am}^{\alpha /2}}} + {\textbf{N}_m}},&{{\Psi_0}},\\
	{\frac{{\sqrt {{p_{ab}}} {h_{am}}{\boldsymbol{x}_b}}}{{d_{am}^{\alpha /2}}} + \frac{{\sqrt {{p_{ac}}} {h_{am}}{\boldsymbol{x}_c}}}{{d_{am}^{\alpha /2}}} + {\textbf{N}_m}},&{{\Psi_1}},
	\end{array}} \right.
\end{align}
where $p_{ab}$ and $p_{ac}$ are Alice's transmit power for Bob and Carol, respectively, $\alpha$ is the path-loss exponent, and ${\textbf{N}_m} \sim \mathcal{CN}\left( {\textbf{0},\sigma _m^2 \textbf{I}_n} \right)$
represents the receiver noise  at $m$. Here, $\textbf{I}_n$ represents an $n\times n$ identity matrix. Then notation $\Psi_0$ states that Alice does not transmit a covert signal to Carol, while $\Psi_1$ states that Alice transmits to Carol.
In the following, we assume the total transmit power is limited by $P$, which is a common assumption in the literature \cite{STPC0}--\cite{wong}. Hence, Alice transmits secure and covert messages (to Bob and Carol)  with power ${p_{ab}} = \left\{ {\begin{array}{*{20}{l}}
	{{\rho _s}P}&{{\Psi _0}}\\
	{{\rho _{cs}}P}&{{\Psi _1}}
	\end{array}} \right.$ and
 $
 {p_{ac}} = \left\{ {\begin{array}{*{20}{l}}
 	0&{{\Psi _0}}\\
 	{\left( {1 - {\rho _{cs}}} \right)P}&{{\Psi _1}}
 	\end{array}} \right.
 $, respectively, where $\rho_s \in [0,1]$ and $\rho_{cs} \in [0,1]$ are the power
allocation factor in $\Psi_0$ and $\Psi_1$ slots, respectively. 
In order to simplify notations, we define ${\gamma _c} = \frac{{P{{\left| {{h_{ac}}} \right|}^2}}}{{d_{ac}^\alpha \sigma _c^2}},{\gamma _b} = \frac{{P{{\left| {{h_{ab}}} \right|}^2}}}{{d_{ab}^\alpha \sigma _b^2}},{\gamma _{u}} = \frac{{P{{\left| {{h_{au}}} \right|}^2}}}{{d_{au}^\alpha \sigma _{u}^2}},{\gamma _{w}} = \frac{{P{{\left| {{h_{aw}}} \right|}^2}}}{{d_{aw}^\alpha \sigma _{w}^2}}$.
The signal-to-noise ratio (SNR) and the signal-to-interference-plus-noise-ratio (SINR) for symbol $\ell$ at the untrusted user and Bob  can be written, respectively,  as follows
\begin{align}
\gamma _{U}^\ell {\rm{ = }}\left\{ {\begin{array}{*{20}{l}}
	{\rho_s {\gamma _{u}},}&{{\Psi _0},}\\
	{}&{}\\
	{\frac{{\rho_{cs} {\gamma _{u}}}}{{1 + \left( {1 - \rho_{cs} } \right){\gamma _{u}}}},}&{{\Psi _1},}
	\end{array}} \right.\,\,\,\,
\end{align}
\begin{align}
\gamma _B^\ell {\rm{ = }}\left\{ {\begin{array}{*{20}{l}}
	{\rho_s {\gamma _b},}&{{\Psi _0},}\\
	{}&{}\\
	{\frac{{\rho_{cs} {\gamma _b}}}{{1 + \left( {1 - \rho_{cs} } \right){\gamma _b}}},}&{{\Psi _1}.}
	\end{array}} \right.
\end{align}
 Therefore, the secrecy rate at Bob is given by
\begin{align}
R_{sec} ^\ell\left( \rho  \right) = {\left[ {\log_2 \left( {1 + \gamma _B^\ell } \right) - \log_2 \left( {1 + \gamma _{U}^\ell } \right)} \right]^ + },
\end{align}
where ${\left[ x \right]^ + }$ is defined as $\max \left\{ {x,0} \right\}$.

\subsection{Covert Communication Requirement}\label{Covert communication}
Based on its received signal power, the warden decides  whether Alice has sent data to Carol or not.
If the warden decides that Alice has sent data to Carol  when Alice has not sent any data to Carol, this means that a false alarm (FA) with probability of $p_r^{FA}$ has occurred. Moreover, if the warden decides that Alice has not sent data to Carol when Alice has sent data to Carol, then we say that a missed detection (MD) with probability of $p_r^{MD}$ has occurred.
Each symbol of the received signal at the warden i.e., $y_{w}^\ell$,  has the following distribution
\begin{align}
y_{w}^\ell  \sim  {\cal C}{\cal N}\left( {0,\sigma _{w}^2 + \Im  } \right),
\end{align}
where $\Im  = \left\{ {\begin{array}{*{20}{l}}
	{\frac{{{\rho _s}P}}{{d_{aw}^\alpha }}{{\left| {{h_{aw}}} \right|}^2}},&{{\Psi _0}},\\\\
	{\frac{\rho_{cs} P+\left( {1 - {\rho _{cs}}} \right)P}{d_{aw}^\alpha}{{\left| {{h_{aw}}} \right|}^2}},&{{\Psi _1}},
	\end{array}} \right.$ and  the probability density function (PDF) of  $\Im $ is  \cite{jammer}
\begin{align}\label{distribution}
{f _\Psi}\left( \Im   \right) = \left\{ {\begin{array}{*{20}{l}}
	{\frac{1}{{{\psi _0}}}{e^{ - \frac{\Im  }{{{\psi _0}}}}}},&{\Im   > 0,\,\,\,{\Psi_0}},\\
	{\frac{1}{{{\psi _1}}}{e^{ - \frac{\Im  }{{{\psi _1}}}}}},&{\Im   > 0,\,\,\,{\Psi_1}},
	\end{array}} \right. 
\end{align}
where 
$\psi_0= \frac{\rho_s P}{d_{aw}^\alpha}$ and $\psi_1= \frac{\rho_{cs} P+\left( {1 - {\rho _{cs}}} \right)P}{d_{aw}^\alpha}= \frac{P}{d_{aw}^\alpha}$. The received SINR for symbol $\ell$ at Carol is given by
\begin{align}
\gamma_C^\ell{\rm{ = }}\left\{ {\begin{array}{*{20}{l}}
	{0,}&{{\Psi _0},}\\
	{\frac{{\left( {1 - \rho_{cs} } \right){\gamma _c}}}{{1 + \rho_{cs} {\gamma _c}}},}&{{\Psi _1}.}
	\end{array}} \right.
\end{align} 
Note that Alice has successfully achieved covert communication with Carol when the following inequality  is satisfied \cite{jammer}
\begin{align}\label{decision}
\text{for any}\, \varepsilon \ge 0, \,\,\, p_r^{MD}+p_r^{FA}\ge 1-\varepsilon, \,\,\, \text{as}\,\,\, n \to \infty.
\end{align}
Moreover, the optimal decision rule  for minimizing the detection error at the warden is written as  \cite{jammer}
\begin{align}
\frac{{{Y_{w}}}}{n}	\mathop \gtrless\limits_{\Psi_0}^{\Psi_1} \theta,
\end{align}
where ${Y_{w}} = \sum\limits_{\ell  = 1}^n {{{\left| {y_{w}^\ell } \right|}^2}} $
is the total received power at the warden in each time slot and $\theta $ is the decision threshold at the warden. The FA and MD  probabilities can be written as
\begin{align}\label{PFA}
{p_r^{FA}} = \mathbb{P}\left( {\frac{{{Y_{w}}}}{n} > \theta \left| {{\Psi_0}} \right.} \right) = \mathbb{P}\left( {\left( {\sigma _{w}^2 + \Im  } \right)\frac{{\chi _{2n}^2}}{n} > \theta \left| {{\Psi_0}} \right.} \right),
\end{align}
\begin{align}\label{PMD}
{p_r^{MD}} = \mathbb{P}\left( {\frac{{{Y_{w}}}}{n} < \theta \left| {{\Psi_1}} \right.} \right) = \mathbb{P}\left( {\left( {\sigma _{w}^2 + \Im  } \right)\frac{{\chi _{2n}^2}}{n} < \theta \left| {{\Psi_1}} \right.} \right),
\end{align}
where ${\chi _{2n}^2}$ is a chi-squared random variable with $2n$ degrees of freedom. 
According to the Strong Law of Large Numbers (SLLN),
$\frac{{\chi _{2n}^2}}{n}$  converges to 1, and based on 
Lebesgue’s Dominated Convergence Theorem \cite{coverage}
we can replace  $\frac{{\chi _{2n}^2}}{n}$  with 1, when $n \to \infty $ . Hence we can rewrite \eqref{PFA} and \eqref{PMD}  as follows

\begin{align}\label{pfa}
	{p_r^{FA}} = \mathbb{P}\left( { {\sigma _{w}^2 + \Im  } > \theta \left| {{\Psi_0}} \right.} \right),
\end{align}
\begin{align}\label{pmd}
	{p_r^{MD}}  = \mathbb{P}\left( { {\sigma _{w}^2 + \Im  } < \theta \left| {{\Psi_1}} \right.} \right),
\end{align}
by using  distribution of random variable $\Im $ as explained in  \eqref{distribution}, \eqref{pfa} and \eqref{pmd} are  calculated as follows
\begin{align}\label{FA_C}
{p_r^{FA}}{\rm{ = }}\left\{ {\begin{array}{*{20}{l}}
	{{e^{ - \frac{{\left( {\theta  - \sigma _{w}^2} \right)}}{{{\psi _0}}}}},}&{\theta  - \sigma _{w}^2 \ge 0,}\\
	{1,}&{\theta  - \sigma _w^2 < 0,}
	\end{array}} \right.
\end{align}
\begin{align}\label{MD_C}
{p_r^{MD}}{\rm{ = }}\left\{ {\begin{array}{*{20}{l}}
	{1 - {e^{ - \frac{{\left( {\theta  - \sigma _{w}^2} \right)}}{{{\psi _1}}}}},}&{\theta  - \sigma _{w}^2 \ge 0,}\\
	{0,}&{\theta  - \sigma _{w}^2 < 0.}
	\end{array}} \right.
\end{align}

By exploiting \eqref{FA_C} and \eqref{MD_C}, ${p_r^{FA}} + {p_r^{MD}}$ can be written as
\begin{align}\label{MD-FA0}
{p_r^{FA}} + {p_r^{MD}}{\rm{ = }}\left\{ {\begin{array}{*{20}{l}}
	{1 - {e^{ - \frac{{\left( {\theta  - \sigma _{w}^2} \right)}}{{{\psi _1}}}}} + {e^{ - \frac{{\left( {\theta  - \sigma _{w}^2} \right)}}{{{\psi _0}}}}},}&{\theta  - \sigma _{w}^2 \ge 0,}\\
	{1,}&{\theta  - \sigma _{w}^2 < 0.}
	\end{array}} \right.
\end{align}
It is clear that the warden will select a decision threshold greater than the  variance of the received noise, to ensure that the resulting detection error probability will be less than $1$.

\subsection{Optimization Formulation}\label{Optimization Problem}
In this section, to evaluate the proposed system model we formulate a novel optimization problem at which the main aim is to maximize the average rate subject to transmit power limitation,  quality of service constraints, and  the covert communication requirement i.e.,  \eqref{decision}.
The secrecy rate at Bob is given by
\begin{align}\label{sr}
&{R_{\sec }^\ell }\left(  \rho_s, \rho_{cs} \right) =\nonumber\\&\hspace{-.1cm} \left\{ {\begin{array}{*{20}{l}}
	\hspace{-.2cm}
	{{{\left[ {\log_2 \left( {1 + {\rho_s \gamma _b}} \right) - \log_2 \left( {1 + {\rho_s \gamma _u}} \right)} \right]}^ + },}&\hspace{-.2cm}{{\Psi _0},}\\
	{}&{}\\
	\hspace{-.2cm}
	{{{\left[ {\log_2 \left( {1 + \frac{{\rho_{cs}{\gamma _b}}}{{1 + (1-\rho_{cs}) {\gamma _b}}}} \right) - \log_2 \left( {1 + \frac{{\rho_{cs}{\gamma _u}}}{{1 + (1-\rho_{cs}) {\gamma _u}}}} \right)} \right]}^ + },}&\hspace{-.2cm}{{\Psi _1}.}
	\end{array}} \right. 
\end{align}
When Alice transmits a secure message and covert message to Bob and Carol, respectively, the sum rate in  this time slot is
\begin{align}
\begin{array}{l}
R = \log_2 \left( {1 + \frac{{(1-\rho_{cs}){\gamma _c}}}{{1 +  \rho_{cs} {\gamma _c}}}} \right)+\\
\,\,\,\,\,\,  {\left[ {\log_2 \left( {1 + \frac{{\rho_{cs}{\gamma _b}}}{{1 + (1-\rho_{cs}) {\gamma _b}}}} \right) - \log_2 \left( {1 + \frac{{\rho_{cs}{\gamma _u}}}{{1 + (1-\rho_{cs}) {\gamma _u}}}} \right)} \right]^ + }.
\end{array}
\end{align}
Since Alice does not transmit data to Carol continuously, we define two parameters $r_0$ and $r_1$, such that $r_0= 1-r_1$. When Alice only transmits a secure message to Bob, we set  $r_0=1$.  When Alice transmits both a secure message and covert message, we set  $r_1=1$. Hence, the sum rate in each time slot can be expressed as
\begin{align}
&\hat R = {r_0}{\left[ {{{\log }_2}\left( {1 + {\rho _s}{\gamma _b}} \right) - {{\log }_2}\left( {1 + {\rho _s}{\gamma _u}} \right)} \right]^ + } + {r_1}\left[ {{{\log }_2}\left( 1+ \right.} \right. \nonumber\\&
\left. {  \frac{{{\rho _{cs}}{\gamma _b}}}{{1 + (1 - {\rho _{cs}}){\gamma _b}}}} \right){\left. { - {{\log }_2}\left( {1 + \frac{{{\rho _{cs}}{\gamma _u}}}{{1 + (1 - {\rho _{cs}}){\gamma _u}}}} \right)} \right]^ + } + \nonumber\\&
{r_1}{\log _2}\left( {1 + \frac{{(1 - {\rho _{cs}}){\gamma _c}}}{{1 + {\rho _{cs}}{\gamma _c}}}} \right).
\end{align}
We assume  $r_1=1$ and $r_0=1$ occur with probability of $p_r^{r_1}$ and $p_r^{r_0}$, respectively, where $p_r^{r_0}+p_r^{r_1}=1$.
Consequently, the average  rate in this network can be written as
\begin{align}\label{avr}
&\bar R ={\rm E_{r_0,r_1}}\left\{ {{\hat R} } \right\} =\nonumber\\& p_r^{{r_0}}{\left[ {\log_2 \left( {1 +\rho_s {\gamma _b}} \right) - \log_2 \left( {1 + \rho_s{\gamma _u}} \right)} \right]^ + } + p_r^{{r_1}}\left[ {{{\log }_2}\left( 1+ \right.} \right. \nonumber\\&
\left. {  \frac{{{\rho _{cs}}{\gamma _b}}}{{1 + (1 - {\rho _{cs}}){\gamma _b}}}} \right){\left. { - {{\log }_2}\left( {1 + \frac{{{\rho _{cs}}{\gamma _u}}}{{1 + (1 - {\rho _{cs}}){\gamma _u}}}} \right)} \right]^ + } \nonumber\\&+ p_r^{{r_1}}\log_2 \left( {1 + \frac{{(1-\rho_{cs}){\gamma _c}}}{{1 +  \rho_{cs} {\gamma _c}}}} \right),
\end{align}
where ${\rm E_{r_0,r_1}}\left\{ {.} \right\}$ is the expectation operator with respect to random variables $r_0$ and $r_1$.
In order to maximize the average rate subject to the power limitation, covert communication requirement, and  secrecy rate constraints, we propose the following optimization problem
\begin{subequations}\label{Opt}
	\begin{align}
&{\mathop {\max }\limits_{  \rho_{s}, \rho_{cs}}  \;\bar R\left(  \rho_{s}, \rho_{cs}  \right),}\\&
{{\rm{s}}.{\rm{t}}.:0 \le \rho_{s}  \le 1} \label{rhos}\\&
\hspace{.8cm}0\le \rho_{cs}  \le 1\label{rhocs}\\&\label{QoSs}
{{\kern 1pt} {\kern 1pt} {\kern 1pt} {\kern 1pt} {\kern 1pt} {\kern 1pt} {\kern 1pt} {\kern 1pt} {\kern 1pt} {\kern 1pt} {\kern 1pt} {\kern 1pt} {\kern 1pt} \,\,\,\,\,{\kern 1pt} {\kern 1pt} p_r^{{r_0}}{{[\log_2(1 + \rho_s{\gamma _b}) - \log_2(1 + \rho_s {\gamma _u})]}^ + }}\\&\nonumber
{{\kern 1pt} {\kern 1pt} {\kern 1pt} {\kern 1pt} {\kern 1pt} {\kern 1pt} {\kern 1pt} {\kern 1pt} {\kern 1pt} {\kern 1pt} {\kern 1pt} {\kern 1pt} {\kern 1pt} {\kern 1pt} {\kern 1pt} {\kern 1pt} {\kern 1pt} {\kern 1pt}  + p_r^{{r_1}}\left[ {\log_2 \left( {1 + \frac{{\rho_{cs}{\gamma _b}}}{{1 + (1-\rho_{cs}) {\gamma _b}}}} \right)} \right.}\\&\nonumber
{{\kern 1pt} {\kern 1pt} {\kern 1pt} {\kern 1pt} {\kern 1pt} {\kern 1pt} {\kern 1pt} {\kern 1pt} {\kern 1pt} {\kern 1pt} {\kern 1pt} {\kern 1pt} {\kern 1pt} {\kern 1pt} {\kern 1pt} {\kern 1pt} {\kern 1pt} {\kern 1pt} {{\left. { - \log_2 \left( {1 + \frac{{\rho_{cs}{\gamma _u}}}{{1 +(1- \rho_{cs}) {\gamma _u}}}} \right)} \right]}^ + } \ge {R_{\sec }^{\min }},}\\&\label{QoSc}
{{\kern 1pt} {\kern 1pt} {\kern 1pt} {\kern 1pt} {\kern 1pt} {\kern 1pt} \,\,{\kern 1pt} {\kern 1pt} {\kern 1pt} {\kern 1pt} {\kern 1pt} {\kern 1pt} {\kern 1pt} \,\,\,{\kern 1pt} {\kern 1pt} p_r^{{r_1}}\log_2 \left( {1 + \frac{{(1-\rho_{cs}){\gamma _c}}}{{1 + \rho_{cs} {\gamma _c}}}} \right) \ge R_{{\rm{cov}}}^{\min }.}\\& \label{cc}
\hspace{.8cm}
{\mathop {\min }\limits_{  \vartheta}  \ (p_r^{MD}+p_r^{FA})\ge 1-\varepsilon.} 
\end{align}
\end{subequations}
Constraints \eqref{QoSs} and \eqref{QoSc} are the secure and  covert communication rate requirements, respectively.  Constraint \eqref{cc} is the worst case requirement for the  covert communication at Carol.
\subsection{Proposed Optimization Solution}\label{solution of optimization problem}
To solve the optimization in \eqref{Opt}, we present two lemmas as follows.

	\begin{lemma}\label{L1}
		The optimal power allocation factor in the $\Psi_0$ slot is equal to one i.e., $\rho_s=1$.
	\end{lemma}
	\textit{Proof}:
	The secrecy rate which is defined in \eqref{sr}, is an
	increasing function with respect to $\rho_s$ for $\gamma_b > \gamma_u$. Hence, in order to maximize the average rate, Alice should transmit secure data ($\boldsymbol{x}_b$)  with the maximum allowable transmit power i.e., $P$ in the time slot $\Psi_0$, which leads to $\rho_s=1$. \hspace{1cm} $\blacksquare$
	\begin{lemma}\label{L2}
		According to Lemma \ref{L1}, the covert communication requirement is always satisfied.
	\end{lemma}
	\textit{Proof}:
	When $\rho_s=1$, we have the following equation
	\begin{align}\label{equal}
	\psi_0= \frac{P}{d_{aw}^\alpha}=\psi_1=\psi,
	\end{align}	
	by substituting  \eqref{equal} into  \eqref{MD-FA0} we have
	\begin{align}
	{p_r^{FA}} + {p_r^{MD}}=&\left\{ {\begin{array}{*{20}{l}}
		{1 - {e^{ - \frac{{\left( {\theta  - \sigma _{w}^2} \right)}}{{{\psi}}}}} + {e^{ - \frac{{\left( {\theta  - \sigma _{w}^2} \right)}}{{{\psi }}}}},}&{\theta  - \sigma _{w}^2 \ge 0,}\nonumber\\
		{1,}&{\theta  - \sigma _{w}^2 < 0,}
		\end{array}} \right.\\ \label{e1}
	=& \left\{ {\begin{array}{*{20}{l}}
		{1,}&{\theta  - \sigma _{w}^2 \ge 0,}\\
		{1,}&{\theta  - \sigma _{w}^2 < 0,}
		\end{array}} \right.
	\end{align}
	which satisfies \eqref{cc}. \hspace{4.9cm}$\blacksquare$

The objective function in \eqref{Opt} is non-concave, therefore, convex optimization methods cannot be directly applied to solve the optimization. Hence, we proceed by applying the epigraph method \cite{Boyd}, such that the optimization problem can be rewritten as
\begin{subequations}\label{Opt1}
	\begin{align}
&\mathop {\max }\limits_{ \rho_{cs}, \eta}   \;p_r^{{r_0}}[{\log_2 \left( {1 + {\gamma _b}} \right) - \log_2 \left( {1 + {\gamma _u}} \right)}]^+  \\& \nonumber
\,\,\,\,\,\,\,\,\,\,\,\,\,+ p_r^{{r_1}}\eta  + p_r^{{r_1}}\log_2 \left( {1 + \frac{{(1-\rho_{cs}){\gamma _c}}}{{1 +\rho_{cs} {\gamma _c}}}} \right)
\\&
{\rm{s}}{\rm{.t}}{\rm{.}}:\,\eqref{rhos}, \eqref{rhocs}, \eqref{QoSc},\nonumber
\\&
\hspace{.9cm} p_r^{{r_0}}{\left[\log_2\left(\frac{1 + {\gamma _b}}{1 +  {\gamma _u}}\right) \right]}^ + + p_r^{r_1}\eta \ge {R_{\sec }^{\min }}, \label{Qos_epi}
\\&
\hspace{.9cm} \log_2 \left( {1 + \frac{{ \rho_{cs}{\gamma _b}}}{{1 +(1-\rho_{cs}) {\gamma _b}}}} \right)\label{epic}\\&
\hspace{1.5cm} - \log_2 \left( {1 + \frac{{ \rho_{cs}{\gamma _{u}}}}{{1 +(1-\rho_{cs}){\gamma _{u}}}}} \right) \le \eta, \nonumber\\&
\hspace{.9cm}\eta  \ge 0.
\end{align}
\end{subequations}
The optimization problem \eqref{Opt1} is still non-convex due to constraints \eqref{QoSc} and  \eqref{epic} and the objective function.
To tackle this
non-convexity, we employ the successive convex approximation method to approximate the objective function and constraint \eqref{QoSc} to concave functions and 
constraint  \eqref{epic}  to  a convex constraint. First we consider the objective function. which can be re-expressed as
\begin{align}
\Xi \left( \rho_{cs}  \right) =\Phi \left( \rho_{cs}  \right) - \Gamma \left( \rho_{cs}  \right),
\end{align}
where 
\begin{align}
\left\{ \begin{array}{l}
\Phi \left( \rho_{cs}  \right) = p_r^{{r_0}} [{\log_2 \left( {1 + {\gamma _b}} \right) - \log_2 \left( {1 + {\gamma _u}} \right)}] \\
\,\,\,\,\,\,\,\,\,\,\,\,\,\,\,\,  + p_r^{{r_1}}\eta  + p_r^{{r_1}}\log_2 \left( {1 +  {\gamma _c}} \right),\\
\Gamma \left( \rho_{cs}  \right) = p_r^{{r_1}}\log_2 \left( {1 + \rho_{cs} {\gamma _c}} \right).
\end{array} \right.
\end{align}
Employing the difference of convex
functions (DC) method, we approximate $\Gamma \left( \rho_{cs}  \right) $ as
\begin{align}\label{tilde}
& \Gamma \left( \rho_{cs}  \right) \simeq \tilde \Gamma \left( \rho_{cs}  \right)=	\Gamma\left( {\rho_{cs} \left( {\mu  - 1} \right)} \right) + \\&{\nabla}\Gamma\left( {{\rho_{cs}}\left( {\mu  - 1} \right)} \right)\nonumber\left( {{\rho_{cs}} - {\rho_{cs}}(\mu  - 1)} \right),
\end{align}
where ${\nabla}$ is the gradient operator, $\mu$ is the iteration number, and ${\nabla}\Gamma\left( {{\rho_{cs}}\left( {\mu  - 1} \right)} \right)$ is calculated as
\begin{align}
{\nabla }\Gamma \left( {\rho_{cs} \left( {\mu  - 1} \right)} \right) = \frac{{p_r^{{r_1}}}}{{\ln 2}}\frac{{ {\gamma _c}}}{{1 +  \rho_{cs} (\mu  - 1){\gamma _c}}}.
\end{align}
Finally, the objective function can be rewritten as $\Phi \left( \rho_{cs}  \right) - \tilde \Gamma \left( \rho_{cs}  \right),$ which is concave. 
Similar to the objective function, we can approximate  \eqref{QoSc} and  \eqref{epic} as
 ${\rm T}(\rho_{cs} ) - \tilde\Lambda (\rho_{cs} ) \ge 0$ and
$\tilde \Omega \left( \rho_{cs}  \right) -  \Sigma \left( \rho_{cs}  \right) \le 0$,  respectively, 
where 
${\rm T}(\rho_{cs} ) = p_r^{{r_1}}\log_2 (1 + {\gamma _c}) - R_{{\mathop{\rm cov}} }^{\min }$,
$\Lambda (\rho_{cs} ) = p_r^{{r_1}}\log_2 (1 + \rho_{cs} {\gamma _c})$,
$\Omega (\rho_{cs} ) = \log_2(1 + (1 - \rho_{cs} ){\gamma _u})$, and $\Sigma (\rho_{cs} ) = \log_2(1 + (1 - \rho_{cs} ){\gamma _b}) + \log_2(1 + {\gamma _u}) - \log_2(1 + {\gamma _b}) + \eta $.
Moreover,  $\tilde\Lambda$ and $\tilde \Omega$ can be evaluated similar to \eqref{tilde}. Therefore, after applying the DC approximation,  \eqref{Opt1} can be  rewritten as follows
	\begin{align}\label{opt_epi}
	&{\mathop {\max }\limits_{ \rho_{cs}, \eta}   \;\Phi \left( \rho_{cs}  \right) - \tilde \Gamma \left( \rho_{cs}  \right),}\\&\nonumber
	{{\rm{s}}.{\rm{t}}:\,\eqref{rhos}, \eqref{rhocs}, \eqref{Qos_epi} }
	\\& \nonumber\hspace{.7cm}
	{\rm{T}}(\rho_{cs} ) - \tilde \Lambda (\rho_{cs} ) \ge 0,
	\\&\nonumber\hspace{.7cm}
	\tilde \Omega \left( \rho_{cs}  \right) - \Sigma \left( \rho_{cs}  \right) \le 0.
	\end{align}
Now, optimization problem \eqref{opt_epi} is convex and can be solved using numerical software such as CVX \cite{CVX}.
\begin{algorithm}[t]
	\caption{Iterative Power Allocation Algorithm} \label{alg1}
	\begin{algorithmic}[1]
		\STATE  \nonumber
		Initialization: Set $\mu =0
		\left( {\mu \text{\hspace{.2cm}is the iteration number}} \right)$
		and initialize to $\rho_{cs}(0)$ .
		\STATE \label{set}
		Set $\rho_{cs}=\rho_{cs}(\mu)$,
		\STATE  		
		Solve \eqref{opt_epi} and set  the result  to $\rho_{cs}(\mu +1)$
		\STATE 
		If $\left| {\rho_{cs}\left( {\mu  + 1} \right) - \rho_{cs} \left( \mu  \right)} \right| \le \vartheta  $,\\
		stop,\\
		else\\
		set $\mu = \mu + 1$ and go back to step \ref{set}.
	\end{algorithmic}
\end{algorithm}
Moreover, since we use the DC method, we need to employ an iterative algorithm as shown in Algorithm \ref{alg1}. This algorithm approaches the optimal solution when the stopping condition, i.e, $\left| {\rho_{cs}\left( {\mu  + 1} \right) - \rho_{cs} \left( \mu  \right)} \right| \le \vartheta  $ is satisfied, where $\vartheta$ is the stopping threshold.

\begin{remark}[]
For the case of $\gamma_u\geq \gamma_b$, it is typically expected that when the wiretap channel is better than the main channel, Alice should stop the transmission of secure message. However, we note that if Alice does not transmit any signal in a time slot, the covert communication requirement is not satisfied. To tackle this issue, we propose that Alice transmits an artificial noise instead of Bob's data to satisfy the covert communication requirement. In this case, the average rate in equation \eqref{avr} can be rewritten as follows
\begin{align}\label{avr1}
&\bar R = p_r^{{r_1}}\log_2 \left( {1 + \frac{{(1-\rho_{cs}){\gamma _c}}}{{1 +  \rho_{cs} {\gamma _c}}}} \right).
\end{align}
As such, the power allocation optimization problem can be expressed as
	\begin{align}\label{Opt.ugb}
	&\mathop {\max }\limits_{ \rho_{cs}}   \;p_r^{{r_1}}\log_2 \left( {1 + \frac{{(1-\rho_{cs}){\gamma _c}}}{{1 +  \rho_{cs} {\gamma _c}}}} \right)
	\\&
	{\rm{s}}{\rm{.t}}{\rm{.}}:\,\eqref{rhos}, \eqref{rhocs}, \eqref{QoSc}.\nonumber
	\end{align}
This optimization problem can be solved similar to the optimization problem in \eqref{Opt1} which is skipped for brevity. 
\end{remark}

\subsection{Special Case: Covert Strategy Known to Carol and Bob}\label{PDNOMA section}
In this subsection, we consider that both Carol and Bob know the covert strategy i.e., they  both have access to Alice's pre-shared secret. The pre-shared secret enables Bob and Carol to know which time slot will be used to transmit the covert message. In this scenario, assuming that Carol and Bob share the same transmission bandwidth, 
 we can employ a PD-NOMA multiple access method in which Bob and Carol can perform successive interference cancellation (SIC). By considering SIC, the received SINRs at Bob, Carol and untrusted user are respectively, given by
\begin{align}
&\gamma _{B, SIC}^\ell {\rm{ = }}\left\{ {\begin{array}{*{20}{l}}
	{\rho_s {\gamma _b},}&{{\Psi _0},}\\
	{}&{}\\
	{\frac{{\rho_{cs} {\gamma _b}}}{{1 + a\left( {1 - \rho_{cs} } \right){\gamma _b}}},}&{{\Psi _1}.}
	\end{array}} \right.\\
&
\gamma _{C, SIC}^\ell {\rm{ = }}\left\{ {\begin{array}{*{20}{l}}
	{0,}&{{\Psi _0},}
	{}&{}\\
	{ { \frac{{(1-\rho_{cs}){\gamma _c}}}{{1 +  (1-a)\rho_{cs} {\gamma _c}}}},}&{{\Psi _1}.}
	\end{array}} \right.\\
\end{align}
 where $a$ represents the condition of SIC implementation i.e., for $\frac{{{{\left| {{h_{ab}}} \right|}^2}}}{{d_{ab}^\alpha}} <\frac{{{{\left| {{h_{ac}}} \right|}^2}}}{{d_{ac}^\alpha}} $, we set $a=1$ and otherwise, $a=0$. 
In this case, the average  rate can be written as
 \begin{align}\label{avr_NOMA}
 &\bar R_{SIC} =\nonumber\\& p_r^{{r_0}}{\left[ {\log_2 \left( {1 +\rho_s {\gamma _b}} \right) - \log_2 \left( {1 + \rho_s{\gamma _u}} \right)} \right]^ + } + p_r^{{r_1}}\left[ {{{\log }_2}\left( 1+ \right.} \right. \nonumber\\&
 \left. {  \frac{{{\rho _{cs}}{\gamma _b}}}{{1 + a(1 - {\rho _{cs}}){\gamma _b}}}} \right){\left. { - {{\log }_2}\left( {1 + \frac{{{\rho _{cs}}{\gamma _u}}}{{1 + (1 - {\rho _{cs}}){\gamma _u}}}} \right)} \right]^ + } \nonumber\\&+ p_r^{{r_1}}\log_2 \left( {1 + \frac{{(1-\rho_{cs}){\gamma _c}}}{{1 +  (1-a)\rho_{cs} {\gamma _c}}}} \right). 
 \end{align}
 Therefore, to maximize the average rate subject to the power limitation, covert  requirement, secrecy rate constraints, and PD-NOMA transmission, we propose the following optimization problem
 \begin{subequations}\label{Opt_SIC}
 	\begin{align}
 	&{\mathop {\max }\limits_{ \rho_{cs}}  \;\bar R_{SIC}\left(  \rho_{s}, \rho_{cs}  \right),}\\&
 	{\rm{s}}.{\rm{t}}.:\eqref{rhos}, \eqref{rhocs}\\&\label{QoSs_SIC}
 	{{\kern 1pt} {\kern 1pt} {\kern 1pt} {\kern 1pt} {\kern 1pt} {\kern 1pt} {\kern 1pt} {\kern 1pt} {\kern 1pt} {\kern 1pt} {\kern 1pt} {\kern 1pt} {\kern 1pt} \,\,\,\,\,{\kern 1pt} {\kern 1pt} p_r^{{r_0}}{{[\log_2(1 + \rho_s{\gamma _b}) - \log_2(1 + \rho_s {\gamma _u})]}^ + }}\\&\nonumber
 	{{\kern 1pt} {\kern 1pt} {\kern 1pt} {\kern 1pt} {\kern 1pt} {\kern 1pt} {\kern 1pt} {\kern 1pt} {\kern 1pt} {\kern 1pt} {\kern 1pt} {\kern 1pt} {\kern 1pt} {\kern 1pt} {\kern 1pt} {\kern 1pt} {\kern 1pt} {\kern 1pt}  + p_r^{{r_1}}\left[ {\log_2 \left( {1 + \frac{{\rho_{cs}{\gamma _b}}}{{1 + a(1-\rho_{cs}) {\gamma _b}}}} \right)} \right.}\\&\nonumber
 	{{\kern 1pt} {\kern 1pt} {\kern 1pt} {\kern 1pt} {\kern 1pt} {\kern 1pt} {\kern 1pt} {\kern 1pt} {\kern 1pt} {\kern 1pt} {\kern 1pt} {\kern 1pt} {\kern 1pt} {\kern 1pt} {\kern 1pt} {\kern 1pt} {\kern 1pt} {\kern 1pt} {{\left. { - \log_2 \left( {1 + \frac{{\rho_{cs}{\gamma _u}}}{{1 +(1- \rho_{cs}) {\gamma _u}}}} \right)} \right]}^ + } \ge {R_{\sec }^{\min }},}\\&\label{QoSc_SIC}
 	{{\kern 1pt} {\kern 1pt} {\kern 1pt} {\kern 1pt} {\kern 1pt} {\kern 1pt} \,\,{\kern 1pt} {\kern 1pt} {\kern 1pt} {\kern 1pt} {\kern 1pt} {\kern 1pt} {\kern 1pt} \,\,\,{\kern 1pt} {\kern 1pt} p_r^{{r_1}}\log_2 \left( {1 + \frac{{(1-\rho_{cs}){\gamma _c}}}{{1 + (1-a)\rho_{cs} {\gamma _c}}}} \right) \ge R_{{\rm{cov}}}^{\min }.}\\& \label{cc_SIC}
 	\hspace{.8cm}
 	{\mathop {\min }\limits_{  \vartheta}  \ (p_r^{MD}+p_r^{FA})\ge 1-\varepsilon.} 
 	\end{align}
 \end{subequations}
 In order to solve this optimization problem, we employ the epigraph method \cite{Boyd} and the DC approximation according to Subsection \ref{solution of optimization problem} to convert it to a convex optimization problem. Finally, we can solve the convex optimization by using numerical software such as CVX \cite{CVX}. 

\section{Proposed Optimization with Imperfect Location and Channel State Information}\label{Imperfect informations}
In this section, we consider the more practical scenario where Alice has imperfect knowledge of the warden's location and users' CSI due to the passive warden and channel estimation errors, respectively. 
\subsection{Imperfect Information of Warden's Location}\label{Imperfect Location warden}
In a practical system, Alice will need to estimate the distance between herself and the warden, i.e., $\hat d_{aw}$, but this estimation may have an error defined as  $e_{d_{aw}}= d_{aw}-\hat d_{aw}$ where  $e_{d_{aw}}$ is the estimation error.
We assume that the distance mismatch lies in a bounded set, i.e., 
$\mathbb{E}_{d_{aw}}=\left\{ {{e_{d_{aw}}}: {\left| {{e_{d_{aw}}}} \right|^2} \le  \epsilon_d} \right\}$, where ${\epsilon}_d$ is a known constant.
In this case, the summation of the MD and FA probability is given by:
\begin{align}\label{iml_MD-FA0}
{p_r^{FA}} + {p_r^{MD}}{\rm{ = }}\left\{ {\begin{array}{*{20}{l}}
	{1 - {e^{ - \frac{{\left( {\theta  - \sigma _{w}^2} \right)}}{{{\hat \psi _1}}}}} + {e^{ - \frac{{\left( {\theta  - \sigma _{w}^2} \right)}}{{{\hat \psi _0}}}}},}&{\theta  - \sigma _{w}^2 \ge 0,}\\
	{1,}&{\theta  - \sigma _{w}^2 < 0,}
	\end{array}} \right.
\end{align}
where, $\hat \psi_0= \frac{\rho_s P}{\left(\hat d_{aw}+e_{d_{aw}}\right)^\alpha}$ and $\hat\psi_1= \frac{P}{\left(\hat d_{aw}+e_{d_{aw}}\right)^\alpha}$. In order to maximize the average rate in the imperfect information about warden's  location scenario, we propose the following optimization problem
\begin{subequations}\label{iml_Opt}
	\begin{align}
	&{\mathop {\max }\limits_{  \rho_{s}, \rho_{cs}}  \;\bar R\left(  \rho_{s}, \rho_{cs}  \right),}\\&
	{\rm{s}}.{\rm{t}}.:\eqref{rhos}, \eqref{rhocs}, \eqref{QoSs}, \eqref{QoSc}, \nonumber\\&
 \label{Imp_cc}
	\hspace{.8cm}
	{\mathop {\min }\limits_{  \vartheta,  e_{d_{aw}}}  \ \left(p_r^{MD}+p_r^{FA}\right)\ge 1-\varepsilon,} \\&\hspace{.8cm}
	{\left| {{e_{d_{aw}}}} \right|^2} \le  \epsilon_d.
	\end{align}
\end{subequations}
To solve the optimization problem \eqref{iml_Opt}, we present the following lemma.

\begin{lemma}\label{L3}
	In our joint ITS and covert system model, the optimal power allocation is the same with both perfect and imperfect information of the warden's location. 
\end{lemma}
	\textit{Proof}: According to Lemma \ref{L1}, the optimal power allocation factor in the slot $\Psi_0$ is equal to one i.e., $\rho_s=1$.  
When $\rho_s=1$, we have $\psi_0= \frac{P}{\left(\hat d_{aw}+e_{d_{aw}}\right)^\alpha}=\psi_1=\psi,$
hence, ${p_r^{FA}} + {p_r^{MD}}$ can be written as follows
\begin{align}\label{Imp_MD-FA1}
	{p_r^{FA}} + {p_r^{MD}}=&\left\{ {\begin{array}{*{20}{l}}
			{1 - {e^{ - \frac{{\left( {\theta  - \sigma _{w}^2} \right)}}{{{\psi}}}}} + {e^{ - \frac{{\left( {\theta  - \sigma _{w}^2} \right)}}{{{\psi }}}}},}&{\theta  - \sigma _{w}^2 \ge 0,}\\
			{1,}&{\theta  - \sigma _{w}^2 < 0,}
	\end{array}} \right.\\ \label{Imp_e1}
	=& \left\{ {\begin{array}{*{20}{l}}
			{1,}&{\theta  - \sigma _{w}^2 \ge 0,}\\
			{1,}&{\theta  - \sigma _{w}^2 < 0,}
	\end{array}} \right.
\end{align}
which \eqref{Imp_e1} satisfies \eqref{Imp_cc}. \hspace{4.5cm}$\blacksquare$ 

Finally, to solve \eqref{iml_Opt}, we employ epigraph and DC methods similar to Subsection \ref{solution of optimization problem}.
\subsection{ Imperfect CSI Scenario}\label{IMCSI}
In practical systems, Alice may also have imperfect CSI of the users due to channel estimation errors. Hence, in this subsection, we assume Alice has imperfect CSI of Bob, Carol and the untrusted user. Specifically, Alice has an estimated version of channels \cite{Huang}, \cite{Abedi},  i.e., $\hat h_{ab}$, $\hat h_{ac}$, and $\hat h_{au}$, and the channel estimation errors are
	defined as $e_{h_{ab}} = h_{ab} -\hat h_{ab}$, $e_{h_{ac}} = h_{ac}-\hat h_{ac}$, and  $e_{h_{au}} =  h_{au}-\hat  h_{au}$, respectively. 
	Based on the worst-case method, the channel mismatches lie in the bounded set, i.e., 
	$\mathbb{E}_{h_{ab}}=\left\{ {{e_{h_{ab}}}: {\left| {{e_{h_{ab}}}} \right|^2} \le  \epsilon_b} \right\}$, $\mathbb{E}_{h_{ac}}=\left\{ {{e_{h_{ac}}}: {\left| {{e_{h_{ac}}}} \right|^2} \le  \epsilon_c} \right\}$, and $\mathbb{E}_{h_{au}}=\left\{ {{e_{h_{au}}}: {\left| {{e_{h_{au}}}} \right|^2} \le  \epsilon_u} \right\}$, where ${\epsilon}_b$, ${\epsilon}_c$ and ${\epsilon}_u$ are known constants. Therefore, the channel gains from Alice to the users are modeled as follows:
	\begin{align}
&	{\left|  { h_{ab}} \right|^2} = {\left| {\hat h_{ab} + {e_{h_{ab}}}} \right|^2},
	{\left|  { h_{ac}} \right|^2} =  {\left| {\hat h_{ac} + {e_{h_{ac}}}} \right|^2}, \\& {\left|{ h_{au}} \right|^2} =  {\left| {\hat h_{au} + {e_{h_{au}}}} \right|^2}.
	\end{align}
   In the following, we focus on the worst-case performance, in which we maximize the average rate for the worst channel mismatch $e_{h_{ab}}$, $e_{h_{ac}}$,  and $e_{h_{au}}$ in the bounded set $\mathbb{E}_{h_{ab}}$, $\mathbb{E}_{h_{ac}}$ and $\mathbb{E}_{h_{au}}$, respectively. Hence, the imperfect CSI and imperfect information about warden's  location optimization problem can be formulated as follows: 
	\begin{subequations}\label{Imc_Opt}
		\begin{align}
		&{\mathop {\max }\limits_{  \rho_{s}, \rho_{cs}}  \; \mathop {\hspace{-.2cm}\min }\limits_{{e_{{h_{ab}}}},{e_{{h_{ac}}}},{e_{{h_{au}}}}} \bar R\left(  \rho_{s}, \rho_{cs}  \right),}\\&
		{{\rm{s}}.{\rm{t}}}.:\eqref{rhos}, \eqref{rhocs}\\&\label{Imc_QoSs}
		{{\kern 1pt} {\kern 1pt} {\kern 1pt} {\kern 1pt} {\kern 1pt} {\kern 1pt} {\kern 1pt} {\kern 1pt} {\kern 1pt} {\kern 1pt} {\kern 1pt} {\kern 1pt} {\kern 1pt} \,\,\,\,\,{\kern 1pt} {\kern 1pt} p_r^{{r_0}}{{[\log_2(1 + \rho_s{\gamma _b}) - \log_2(1 + \rho_s {\gamma _u})]}^ + }}\\&\nonumber
		{{\kern 1pt} {\kern 1pt} {\kern 1pt} {\kern 1pt} {\kern 1pt} {\kern 1pt} {\kern 1pt} {\kern 1pt} {\kern 1pt} {\kern 1pt} {\kern 1pt} {\kern 1pt} {\kern 1pt} {\kern 1pt} {\kern 1pt} {\kern 1pt} {\kern 1pt} {\kern 1pt}  + p_r^{{r_1}}\left[ {\log_2 \left( {1 + \frac{{\rho_{cs}{\gamma _b}}}{{1 + (1-\rho_{cs}) {\gamma _b}}}} \right)} \right.}\\&\nonumber
		{{\kern 1pt} {\kern 1pt} {\kern 1pt} {\kern 1pt} {\kern 1pt} {\kern 1pt} {\kern 1pt} {\kern 1pt} {\kern 1pt} {\kern 1pt} {\kern 1pt} {\kern 1pt} {\kern 1pt} {\kern 1pt} {\kern 1pt} {\kern 1pt} {\kern 1pt} {\kern 1pt} {{\left. { - \log_2 \left( {1 + \frac{{\rho_{cs}{\gamma _u}}}{{1 +(1- \rho_{cs}) {\gamma _u}}}} \right)} \right]}^ + } \ge {R_{\sec }^{\min }},}\\&\label{Imc_QoSc}
		{{\kern 1pt} {\kern 1pt} {\kern 1pt} {\kern 1pt} {\kern 1pt} {\kern 1pt} \,\,{\kern 1pt} {\kern 1pt} {\kern 1pt} {\kern 1pt} {\kern 1pt} {\kern 1pt} {\kern 1pt} \,\,\,{\kern 1pt} {\kern 1pt} p_r^{{r_1}}\log_2 \left( {1 + \frac{{(1-\rho_{cs}){\gamma _c}}}{{1 + \rho_{cs} {\gamma _c}}}} \right) \ge R_{{\rm{cov}}}^{\min }.}\\& \label{Imc_cc}
		\hspace{.8cm}
		 {\mathop {\min }\limits_{  \vartheta}  \ \left(p_r^{MD}+p_r^{FA}\right)\ge 1-\varepsilon,}\\&
		\label{Cb}\hspace{.8cm}  {\left| {{e_{h_{ab}}}} \right|^2} \le  \epsilon_b, \\&
		\label{Cc}\hspace{.8cm}  {\left| {{e_{h_{ac}}}} \right|^2} \le  \epsilon_c, \\&
		\label{Cu}\hspace{.8cm}  {\left| {{e_{h_{au}}}} \right|^2} \le  \epsilon_u.
		\end{align}
	\end{subequations}
\subsection{Proposed Optimization Solution}\label{Imc_solution of optimization problem}
In order to solve \eqref{Imc_Opt}, we perform the following two steps: 1) Solving the inner minimization and obtain $e_{h_{ab}}$, $e_{h_{ac}}$, and $e_{h_{au}}$, 2) solving the maximization problem according to Section \ref{solution of optimization problem}. The inner minimization is
formulated as follows:
	\begin{subequations}\label{Imc_im_Opt}
	\begin{align}
	&{ \mathop {\min }\limits_{{e_{{h_{ab}}}},{e_{{h_{ac}}}},{e_{{h_{au}}}}} \bar R\left(  \rho_{s}, \rho_{cs}  \right),}\\&
	{{\rm{s}}.{\rm{t}}}.:\eqref{Imc_QoSs}, \eqref{Imc_QoSc}, \eqref{Cb}-\eqref{Cu}.
	\end{align}
\end{subequations}
In order to solve this optimization problem, we employ the worst-case approach. To this end, 
we employ the triangle inequality which is defined as follows
\begin{align}\label{Tr_enq}
&{\left| {\hat h_{ab}} \right|^2} -\epsilon_b \le {\left| {\hat h_{ab}} \right|^2} - {\left| {{e_{h_{ab}}}} \right|^2} \le {\left| {\hat h_{ab} + {e_{h_{ab}}}} \right|^2} \le \nonumber\\&
{\left| {\hat h_{ab}} \right|^2} + {\left| {{e_{h_{ab}}}} \right|^2} \le {\left| {\hat h_{ab}} \right|^2}+\epsilon_b.
\end{align}
Likewise, we have this inequality for $h_{ac}$ and $h_{au}$. By employing these inequalities, we can write the lower bound of the objective function as  
\begin{align}
&\bar R\left(  \rho_{s}, \rho_{cs}  \right) \ge \bar R_{lb}\left(  \rho_{s}, \rho_{cs}  \right)= \nonumber\\& p_r^{{r_0}}{\left[ {\log_2 \left( {1 +\rho_s {\gamma _b^{lb}}} \right) - \log_2 \left( {1 + \rho_s{\gamma _u^{ub}}} \right)} \right]^ + } + p_r^{{r_1}}\left[ {{{\log }_2}\left( 1+ \right.} \right. \nonumber\\&
\left. {  \frac{{{\rho _{cs}}{\gamma _b^{lb}}}}{{1 + (1 - {\rho _{cs}}){\gamma _b^{ub}}}}} \right){\left. { - {{\log }_2}\left( {1 + \frac{{{\rho _{cs}}{\gamma _u^{ub}}}}{{1 + (1 - {\rho _{cs}}){\gamma _u^{lb}}}}} \right)} \right]^ + } \nonumber\\&+ p_r^{{r_1}}\log_2 \left( {1 + \frac{{(1-\rho_{cs}){\gamma _c^{lb}}}}{{1 +  \rho_{cs} {\gamma _c^{ub}}}}} \right), 
\end{align}
where
\\
$\gamma _c^{lb}=\frac{{P{{\left| {\hat h_{ac}} \right|^2} -\epsilon_c}}}{{d_{ac}^\alpha \sigma _c^2}} \le 
{\gamma _c} = \frac{{P{{\left| {{h_{ac}}} \right|}^2}}}{{d_{ac}^\alpha \sigma _c^2}}\le \gamma _c^{ub}=\frac{{P{{\left| {\hat h_{ac}} \right|^2} +\epsilon_c}}}{{d_{ac}^\alpha \sigma _c^2}}, $
\\
$
\gamma _b^{lb}=\frac{{P{{\left| {\hat h_{ab}} \right|^2} -\epsilon_b}}}{{d_{ab}^\alpha \sigma _b^2}}\le {\gamma _b} = \frac{{P{{\left| {{h_{ab}}} \right|}^2}}}{{d_{ab}^\alpha \sigma _b^2}}\le \gamma _b^{ub}= \frac{{P{{\left| {\hat h_{ab}} \right|^2} +\epsilon_c}}}{{d_{ab}^\alpha \sigma _b^2}},$
\\
$ \gamma _u^{lb}= \frac{{P{{\left| {\hat h_{au}} \right|^2} -\epsilon_b}}}{{d_{au}^\alpha \sigma _{u}^2}} \le {\gamma _{u}} = \frac{{P{{\left| {{h_{au}}} \right|}^2}}}{{d_{au}^\alpha \sigma _{u}^2}}\le  \gamma _u^{ub}=\frac{{P{{\left| {\hat h_{au}} \right|^2} +\epsilon_u}}}{{d_{au}^\alpha \sigma _{u}^2}}$.
\\
\\
Finally, the optimization problem \eqref{Imc_Opt} can be rewritten as
	\begin{subequations}\label{Imcf_Opt}
	\begin{align}
	&{\mathop {\max }\limits_{\rho_{cs}}  \; \bar R_{lb}\left(  \rho_{s}, \rho_{cs}  \right)}\\&
	{{\rm{s}}.{\rm{t}}}.:\eqref{rhos}, \eqref{rhocs}\\&\label{Imcf_QoSs}
	{{\kern 1pt} {\kern 1pt} {\kern 1pt} {\kern 1pt} {\kern 1pt} {\kern 1pt} {\kern 1pt} {\kern 1pt} {\kern 1pt} {\kern 1pt} {\kern 1pt} {\kern 1pt} {\kern 1pt} \,\,\,\,\,{\kern 1pt} {\kern 1pt} p_r^{{r_0}}{{[\log_2(1 + \rho_s{\gamma _b^{lb}}) - \log_2(1 + \rho_s {\gamma _u^{ub}})]}^ + }}\\&\nonumber
	{{\kern 1pt} {\kern 1pt} {\kern 1pt} {\kern 1pt} {\kern 1pt} {\kern 1pt} {\kern 1pt} {\kern 1pt} {\kern 1pt} {\kern 1pt} {\kern 1pt} {\kern 1pt} {\kern 1pt} {\kern 1pt} {\kern 1pt} {\kern 1pt} {\kern 1pt} {\kern 1pt}  + p_r^{{r_1}}\left[ {\log_2 \left( {1 + \frac{{\rho_{cs}{\gamma _b^{lb}}}}{{1 + (1-\rho_{cs}) {\gamma _b^{ub}}}}} \right)} \right.}\\&\nonumber
	{{\kern 1pt} {\kern 1pt} {\kern 1pt} {\kern 1pt} {\kern 1pt} {\kern 1pt} {\kern 1pt} {\kern 1pt} {\kern 1pt} {\kern 1pt} {\kern 1pt} {\kern 1pt} {\kern 1pt} {\kern 1pt} {\kern 1pt} {\kern 1pt} {\kern 1pt} {\kern 1pt} {{\left. { - \log_2 \left( {1 + \frac{{\rho_{cs}{\gamma _u^{ub}}}}{{1 +(1- \rho_{cs}) {\gamma _u^{lb}}}}} \right)} \right]}^ + } \ge {R_{\sec }^{\min }},}\\&\label{Imcf_QoSc}
	{{\kern 1pt} {\kern 1pt} {\kern 1pt} {\kern 1pt} {\kern 1pt} {\kern 1pt} \,\,{\kern 1pt} {\kern 1pt} {\kern 1pt} {\kern 1pt} {\kern 1pt} {\kern 1pt} {\kern 1pt} \,\,\,{\kern 1pt} {\kern 1pt} p_r^{{r_1}}\log_2 \left( {1 + \frac{{(1-\rho_{cs}){\gamma _c^{lb}}}}{{1 + \rho_{cs} {\gamma _c^{ub}}}}} \right) \ge R_{{\rm{cov}}}^{\min }.}\\& \label{Imcf_cc}
	\hspace{.8cm}
	{\mathop {\min }\limits_{  \vartheta}  \ \left(p_r^{MD}+p_r^{FA}\right)\ge 1-\varepsilon}.
	\end{align}
\end{subequations}
Since \eqref{Imcf_Opt} is in the same form as \eqref{Opt}, we can employ epigraph method \cite{Boyd} and DC approximation similar to Subsection \ref{solution of optimization problem} to convert \eqref{Imcf_Opt} to a convex problem that can be solved using CVX.

\section{Numerical Results}\label{Simulation Results}
In this section, we present numerical results to evaluate the performance of  our proposed system model and optimization solution. In the simulations, we assume that Bob requires a higher minimum data rate compared to Carol since the covert requirement typically assumes a low data rate user. If both Bob and Carol require high data rate transmissions, further security strategies such as beamforming \cite{3D} may be employed at Alice.  In the numerical results, the considered simulation parameters are listed in Table \ref{table1}.
\begin{table}[h] 
	\centering
	\caption{ Simulation setting}
	\begin{tabular}{|m{2 em} |m{14 em} |m{6 em}|}  	 
		\hline
		$1-\varepsilon$  & Lower bound
		of detection error probability at warden    & $0.9$ \\
		\hline
		$d_{au}$  & Distance between Alice and untrusted user &   $5$ meters (m)\\ 
		\hline
		$d_{aw}$ & Distance between Alice and warden &  $5$ m   \\
		\hline
		$\alpha$ &   Path-loss exponent & $4$
		\\
		\hline
		$R_{sec}^{\min}$ & Minimum ITS requirement of Bob & $0.5 $ bps/Hz\\
		\hline
		$R_{cov}^{\min}$ & Minimum covert requirement of Carol & $0.1 $ bps/Hz\\
		\hline
		\vspace{.2cm}
		$ p_r^{r_1}$ & Probability of data transmission to Carol & $0.5 $\\
		\hline
	\end{tabular}
	\label{table1}
\end{table}

\begin{figure}[h!]
	\begin{center}
		\includegraphics[width=3.8in,height=3in]{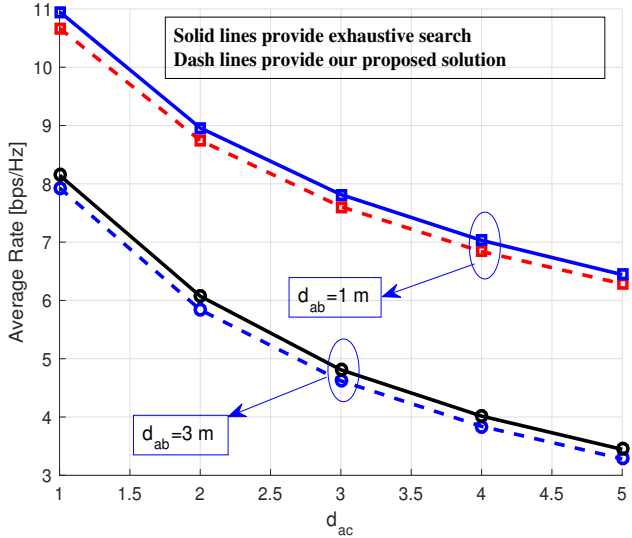}
		\caption{Ergodic average rate \emph{vs}  the distance between Alice and Carol and  $P=3 \,\text{dB}$ , $\sigma _b^2=\sigma_c^2=-33 \,\text{dB}, \sigma _{u}^2=\sigma _{w}^2=-30 \,\text{dB}$.}
		\label{P_dc}
	\end{center}
\end{figure}

Fig. \ref{P_dc}, illustrates the average rate versus the distance between Alice and Carol.  The figure shows that the distance between Alice and Carol has a slightly higher impact on the average rate compared to the distance between Alice and Bob. For example by decreasing $d_{ac}$ from $3$ meters (m) to $1$ m the  average rate increases  on average approximately $39\%$, while by  decreasing $d_{ab}$ from $3$ m to $1$ m the  average rate increases on average approximately $37\%$.when the distance between Alice and Carol is increased, Alice cannot proportionately increase the transmit power $p_{ac}$, due to the covert communication requirement to minimize detection by the warden. Moreover, since Alice transmits data with total available power i.e., $P$ in the secrecy slots $\Psi_0$, decreasing the distance between Alice and Bob $d_{ab}$ has less effect on the average rate.  
Since we apply the DC method to approximate the optimization problem \eqref{Opt1} to a convex one, it is necessary to compare our proposed solution with the optimal solution by employing the exhaustive search method
and finding the optimality gap. As seen in this Fig. \ref{P_dc}, the optimality gap is approximately $5\%$ which highlights the efficiency of our proposed analysis.

\begin{figure}[h!]
	\begin{center}
		\includegraphics[width=3.8in,height=3in]{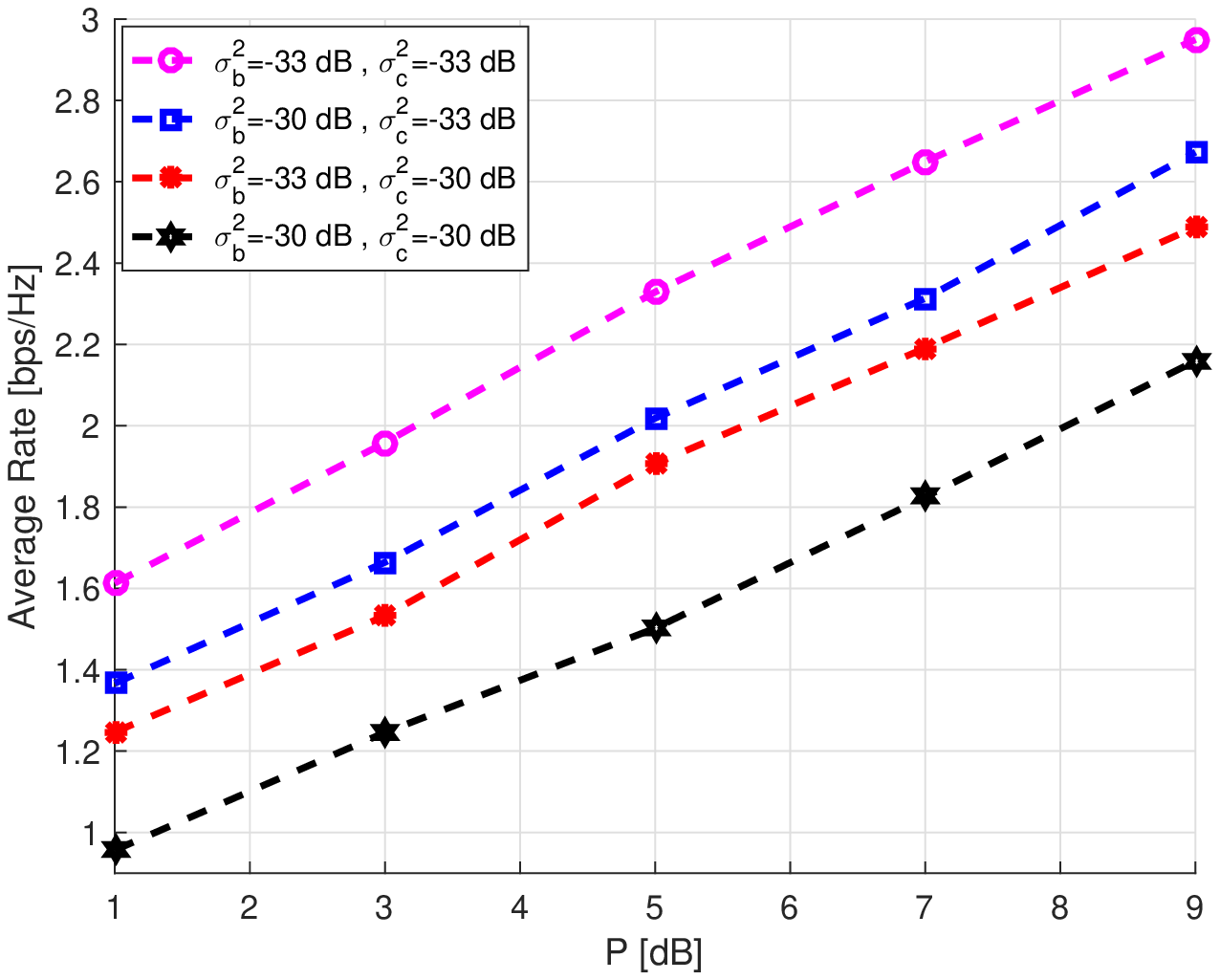}
		\caption{Ergodic average rate \emph{vs}  $P,\ $\, $d_{ac}=d_{ab}=5\, \text{m}$, $\sigma _u^2=\sigma _{w}^2=-30 \,\text{dB}$. }
		\label{pmax_sig_2}
	\end{center}
\end{figure}

Fig. \ref{pmax_sig_2} shows the average  rate versus the total transmit power. As seen in this figure, by increasing  total transmit power, the average rate increases. Moreover this figure shows the impact of the received noise power at Carol on the  average rate is higher than the impact of the received noise power at Bob on the average rate. Similar to Fig. \ref{P_dc}, we see that the covert requirement has a higher impact on the average rate compared to the ITS requirement due to the constraint on Alice increasing $p_{ac}$.



\begin{figure}[t!]
	\begin{center}
		\includegraphics[width=3.8in,height=3in]{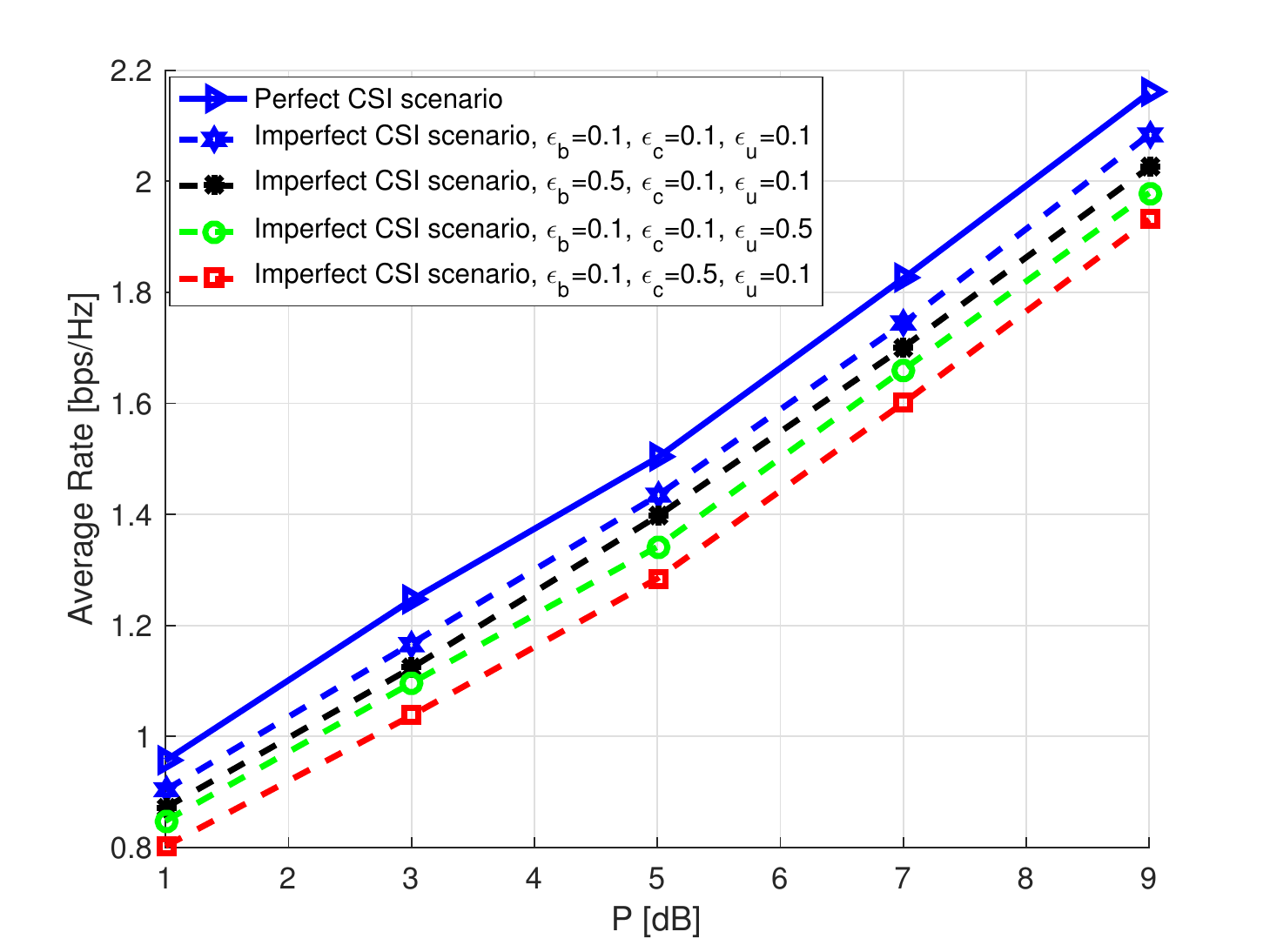}
		\caption{Ergodic average rate \emph{vs}  $P$, $d_{ac}=d_{ab}=5\, \text{m}$, $\sigma _u^2=\sigma _{w}^2=-30 \,\text{dB}$. }
		\label{Imperfect}
	\end{center}
\end{figure}
In Fig. \ref{Imperfect}, we evaluate the impact of the imperfect CSI of users on the average rate. We see in the figure that the imperfect CSI of Carol has a more destructive effect on the average rate with respect to the other users. This means that employing a more accurate estimation of Carol's CSI has a more positive effective on the network performance. The reason is that Alice should select a lower transmit power for Carol's covert message to conceal the signal in the background noise, while Bob's message can be transmitted with a higher power. These facts highlight the sensitivity of Carol's channel estimation error on the achievable rate. Moreover, note that the imperfect CSI scenario (in all three cases) provides lower $\rho_{cs}$. This means that the covert transmit power decreases when the secrecy transmit power increases.


\begin{figure}[h!]
	\begin{center}
		\includegraphics[width=3.8in,height=3in]{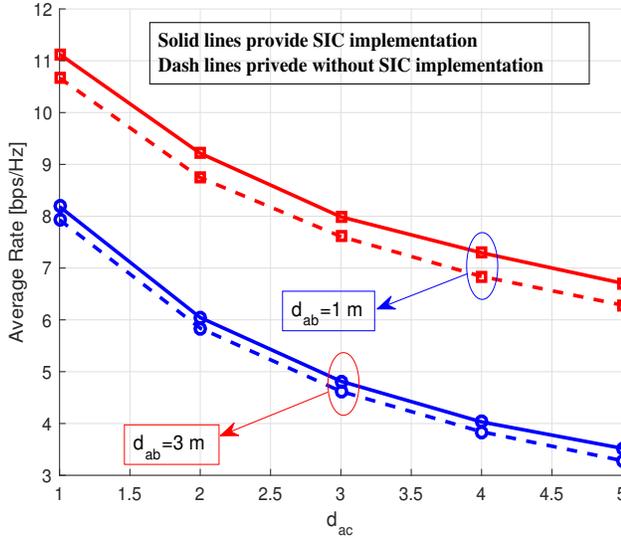}
		\caption{Ergodic average rate \emph{vs}  the distance between Alice and Carol and  $P=3 \,\text{dB}$ , $\sigma _b^2=\sigma_c^2=-33 \,\text{dB}, \sigma _{u}^2=\sigma _{w}^2=-30 \,\text{dB}$.}
		\label{SIC}
	\end{center}
\end{figure}
Fig. \ref{SIC} plots the average rate versus the distance between Alice and Carol.  Moreover, this figure evaluates the impact of SIC implementation on the average rate for the special case when both Carol and Bob know the covert strategy. As seen this figure, when
Carol and Bob know which time slot will be used for the covert message 
and  perform SIC, the average rate is increased by approximately 5\%. As such, the figure shows that there is a fundamental trade-off between the cost of Bob's access to Alice's pre-shared secret  and increase in the average rate. 
\vspace{-.6cm}
\section{Conclusion}\label{Conclusion}

This paper investigates joint ITS and covert communication in a SIMO network, where a source communicates with two legitimate users in the presence of one untrusted user and a warden. 
 While one of the users requests secure communication, the other user needs covert communication. For this system model, we presented an optimization problem with the aim of maximizing the average rate subject to  a covert communication requirement and an ITS rate constraint. To solve the problem, the successive convex approximation method was adopted to convexify the optimization. We then considered a practical system where the location of the warden and the CSIs of users are imperfectly known.
Our numerical examples reveal the impact of network topology and the joint ITS and covert design on the average rate.

\end{document}